# Weather conditions several hours before the strong earthquake

Tao Chen[1], Lei Li[1, 2 *], Xiao-Xin Zhang[3], Qi-Ming Ma[4], Shuo Ti[1], Han Wu[1,2]

[1] State Key Laboratory of Space Weather, National Space Science Center, Chinese Academy of Sciences, Beijing 100190, China

[2] University of Chinese Academy of Sciences, Beijing 100049, China.

[3] Key Laboratory of Space Weather, National Center for Space Weather, China Meteorological Administration, Beijing 100081, China.

[4] Institute of Electrical Engineering, Chinese Academy of Sciences, Beijing 100190, China.

* Correspondence to: Lei Li (2857166529@qq.com)

**Abstract：** Physical phenomena observed before strong earthquake have been reported over centuries. Radon anomalies, electrical signals, water level changes, earthquake lights near the epicenter are recognized as pre-earthquake signals to approach earthquake prediction. Anomalous negative signals observed by ground-based atmospheric electric field instrument under fair weather open up a new way to earthquake prediction. Abnormal heat radiation before the earthquake bring fair weather around the epicenter in theory. In order to figure out the weather conditions around the epicenter before earthquakes, 213 global earthquake events with magnitude of 6 or above from 2013 to 2020 were collected. Based on our definition of fair weather, in 96.7% of the events in the statistics, the weather before the earthquake is fair. Besides, the fair state before the earthquake lasted more than 7 hours, leaving us with enough early warming time.

## 1 Introduction

There are a huge number of reports about strong earthquakes (EQ) all over the world every year. All mankind is eager to be able to accurately warn and notify people before earthquakes in order to take emergency measures to avoid big losses caused by disasters in advance. Therefore, the research of pre-earthquake phenomena is becoming increasingly urgent for scholars around the world. A lot of pre-earthquake phenomena have been summarized and studied by them. For instance, the scholars had found that extremely low frequency electromagnetic waves (ELF, 3 Hz-30 Hz) will appear before strong earthquakes (Dudkin et al., 2013). These waves not only have a strong ability to penetrate the crust, but also decay slowly during propagation. Another advantage of these waves is that they have very little interference. What’s more, there will be various colors of "ground light" before a strong earthquake with different shapes, including flakes, strips, columns and so on. The radon concentration discovered by many scholars was also an important earthquake precursor. (Holub and Bready, 1981; Segovia 1989; King et al. 1993; Pulinets et al., 1997; Garavaglia et al. 2000; Yasuoka et al., 2006; Omori et al., 2007; Kawada et al., 2007).

In addition, atmospheric electric field anomalies before earthquakes are also observed widely.

The early measurements of anomalous atmospheric electric field before earthquakes were made by Bonchkovsky (1954) and Kondo (1968), and recent measurements with modern techniques confirmed the results (Hao, 1988; Smirnov, 2008; Choudhury et al., 2013; Bychkov et al., 2017; Smirnov et al., 2017). Since the atmospheric electric field is abnormal before the earthquake, can we use this method to predict an earthquake? One difficult problem is that, the atmospheric electric field is sensitive to the weather conditions.

Liu et al. (1999) have collected the data of air temperature before Xingtai earthquake in 1966, and the monthly departure of ground temperature in some areas before Tangshan earthquake (May of 1976) to prove thermal omens before earthquakes. He concluded that the temperature will rise obviously before earthquakes. Smirnov et al. (2017) also gave a similar conclusion that temperature increased anomalously before strong earthquakes in Kamchatka region by analyzing the meteorological parameters measured at Paratunka station before three strong earthquakes (M > 7) occurred in Kamchatka in 2006, 2007, and 2016. In this paper, we collect various weather parameters including temperature, relative humidity, wind speed and precipitation, to conduct statistical study on the weather conditions before earthquakes.

## 2 Definition of fair weather

In order to study the weather conditions, it is necessary to have a clear definition of fair days. The definition must have clear quantitative criteria, so as to uniquely determine the data that can be classified as fair days. But the definition is different in various situations. Israelsson and Tammet (2001) classify data with electric field values between 0 and 250 V/m as sunny data. The definition of fair days given by Latha (2003) is that the cloud volume in the sky is less than 3/8, the wind speed is less than 4 m/s and there is no precipitation. The different definition given by Harrison (2004) is defined as no precipitation, no low clouds, the cumulus clouds in the sky is less than 3/8 and the average wind speed is less than 8 m/s. Similar definitions have also been proposed by other scholars. Imyanitov and Shifrin (1962) define fair weather as absence of clouds, precipitation, fog, dust and strong winds. Deshpande and Kamra (2001) considered that fair weather existed when there was no rain or snow, no low clouds, less than 3 oktas of high cloud and the wind speed was less than 10 m/s. Minamoto and Kadokura (2011) gave a criteria of fair day that wind speed of less than 6 m/s and a cloud coverage level of “0” or “0+”. Siingh et al. (2013) also made fair weather data selection by requiring wind speeds less than 10 m/s. Based on these different definitions, we use the following criteria to define fair days: there is no precipitation and average wind speed is less than 8 m/s (28.8 km/h).

The stronger the earthquake, the more obvious the special weather phenomena before it.So, we choose the strong earthquake of magnitude 6 and above all over the world as samples. More than 200 cases from 2013 to 2020 were collected from Global Earthquake Historical Data Query ( http://ditu.92cha.com/dizhen.php ). The weather conditions in several hours before the earthquake are the most meaningful for pre-earthquake warning, so we collected 213 cases of comprehensive statistical weather data 7 hours before the strong earthquake from WEATHER UNDERGROUND ( https://www.wunderground.com ).

## 3 Statistical results

| Date (UTC） | Magnitude (M) | Longitude (°) | Latitude (°) | Position | Humidity % | Wind Speed (km/h) | Preci- -pitation（mm） |
|---|---|---|---|---|---|---|---|
| 2013-03-25 | 6.2 | -90.46 | 14.49 | 6km ENE of Santa Catarina Pinula, Guatemala | 50-68 | 22-22 | 0.0 |
| 2013-04-16 | 7.7 | 62.00 | 28.03 | 83km E of Khash, Iran | 26-48 | 11-26 | 0.0 |
| 2013-04-16 | 6.6 | 142.54 | -3.21 | 23km ESE of Aitape, Papua New Guinea | 75-78 | 0-22 | 0.0 |
| 2013-06-15 | 6.5 | -86.93 | 11.76 | 46km W of Masachapa, Nicaragua | 62-70 | 15-26 | 0.0 |
| 2013-07-02 | 6.1 | 96.67 | 4.65 | 55km S of Bireun, Indonesia | 86-92 | 0-0 | 0.0 |
| 2013-07-17 | 6.0 | -71.74 | -15.66 | 18km W of Chivay, Peru | 33-38 | 4-15 | 0.0 |
| 2013-08-16 | 6.5 | 174.15 | -41.73 | 29km SE of Blenheim, New Zealand | 81-88 | 30-35 | 0.0 |
| 2013-10-12 | 6.6 | 23.25 | 35.51 | 31km W of Platanos, Greece | 60-60 | 0-0 | 0.0 |
| 2013-10-15 | 7.1 | 124.12 | 9.88 | 4km SE of Sagbayan, Philippines | 72-87 | 11-14 | 0.0 |
| 2013-10-30 | 6.2 | -73.40 | -35.31 | 88km W of Constitucion, Chile | 67-82 | 6-11 | 0.0 |
| 2013-10-31 | 6.3 | 121.44 | 23.59 | 46km SSW of Hualian, Taiwan | 73-73 | 4-4 | 0.0 |
| 2014-03-15 | 6.3 | -80.97 | -5.57 | 16km W of Sechura, Peru | 45-58 | 13-15 | 0.0 |
| 2014-03-16 | 6.7 | -70.70 | -19.98 | 64km WNW of Iquique, Chile | 60-73 | 19-33 | 0.0 |
| 2014-03-17 | 6.4 | -70.88 | -20.02 | 80km WNW of Iquique, Chile | 64-68 | 11-20 | 0.0 |
| 2014-03-22 | 6.2 | -70.87 | -19.76 | 91km WNW of Iquique, Chile | 50-64 | 19-35 | 0.0 |
| 2014-03-23 | 6.3 | -70.85 | -19.69 | 94km NW of Iquique, Chile | 50-57 | 31-39 | 0.0 |
| 2014-04-01 | 8.2 | -70.77 | -19.61 | 94km NW of Iquique, Chile | 50-82 | 7-41 | 0.0 |
| 2014-04-01 | 6.9 | -70.95 | -19.89 | 91km WNW of Iquique, Chile | 50-82 | 7-41 | 0.0 |
| 2014-04-03 | 6.5 | -70.58 | -20.31 | 46km WSW of Iquique, Chile | 57-77 | 0-35 | 0.0 |
| 2014-04-03 | 7.7 | -70.49 | -20.57 | 53km SW of Iquique, Chile | 77-77 | 0-9 | 0.0 |
| 2014-04-03 | 6.4 | -70.59 | -20.80 | 78km SW of Iquique, Chile | 77-82 | 0-6 | 0.0 |
| 2014-04-04 | 6.3 | -70.65 | -20.64 | 70km SW of Iquique, Chile | 68-73 | 11-22 | 0.0 |
| 2014-04-11 | 6.2 | -70.65 | -20.66 | 72km SW of Iquique, Chile | 53-60 | 4-17 | 0.0 |
| 2014-04-11 | 6.6 | -85.88 | 11.64 | 15km N of Belen, Nicaragua | 39-49 | 15-22 | 0.0 |
| 2014-04-18 | 7.2 | -100.97 | 17.40 | 33km ESE of Petatlan, Mexico | 66-73 | 0-15 | 0.0 |
| 2014-05-04 | 6.0 | 139.42 | 34.91 | 31km E of Ito, Japan | 43-49 | 7-11 | 0.0 |
| 2014-05-08 | 6.4 | -100.75 | 17.23 | 6km WSW of Tecpan de Galeana, Mexico | 74-83 | 6-11 | 0.0 |
| 2014-05-10 | 6.0 | -100.81 | 17.22 | 14km WSW of Tecpan de Galeana, Mexico | 83-83 | 0-0 | 0.0 |
| 2014-07-07 | 6.9 | -92.46 | 14.72 | 4km W of Puerto Madero, Mexico | 94-100 | 0-11 | 0.0 |
| 2014-07-08 | 6.2 | 168.40 | -17.69 | 9km ENE of Port-Vila, Vanuatu | 74-83 | 0-7 | 0.0 |
| 2014-07-29 | 6.3 | -95.65 | 17.68 | 24km SE of Playa Vicente, Mexico | 56-82 | 4-11 | 0.0 |
| 2014-08-18 | 6.2 | 47.70 | 32.70 | 40km E of Dehloran, Iran | 36-49 | 0-15 | 0.0 |
| 2014-08-18 | 6.0 | 47.70 | 32.58 | 42km ESE of Dehloran, Iran | 12-24 | 19-30 | 0.0 |
| 2014-08-23 | 6.4 | -71.44 | -32.70 | 23km WNW of Hacienda La Calera, Chile | 82-87 | 7-15 | 0.0 |
| 2014-10-14 | 7.3 | -88.12 | 12.53 | 74km S of Intipuca, El Salvador | 100-100 | 2-6 | 0.0 |
| 2014-11-22 | 6.2 | 137.89 | 36.64 | 15km N of Omachi, Japan | 44-70 | 2-7 | 0.0 |
| 2014-12-07 | 6.1 | -91.47 | 13.67 | 61km SSW of Nueva Concepcion, Guatemala | 64-82 | 15-19 | 0.0 |
| 2014-12-10 | 6.1 | 122.45 | 25.54 | 84km ENE of Keelung, Taiwan | 69-83 | 0-15 | 0.0 |
| 2015-01-23 | 6.8 | 168.52 | -17.03 | 80km NNE of Port-Vila, Vanuatu | 94-100 | 4-4 | 0.0 |
| 2015-02-13 | 6.2 | 121.43 | 22.64 | 31km ESE of Taitung City, Taiwan | 47-64 | 7-19 | 0.0 |
| 2015-03-10 | 6.2 | -72.99 | 6.78 | 9km NNE of Aratoca, Colombia | 54-73 | 4-19 | 0.0 |
| 2015-03-18 | 6.2 | -73.52 | -36.12 | 75km NNW of Talcahuano, Chile | 50-64 | 6-19 | 0.0 |
| 2015-03-23 | 6.4 | -69.17 | -18.35 | 45km ESE of Putre, Chile | 83-88 | 2-7 | 0.0 |
| 2015-04-16 | 6.0 | 26.82 | 35.19 | 49km SW of Karpathos, Greece | 37-64 | 24-33 | 0.0 |
| 2015-04-24 | 6.1 | 173.01 | -42.06 | 67km NW of Kaikoura, New Zealand | 78-89 | 7-9 | 0.0 |
| 2015-04-25 | 7.8 | 84.73 | 28.23 | 36km E of Khudi, Nepal | 93-93 | 6-15 | 0.0 |
| 2015-05-12 | 7.3 | 86.07 | 27.81 | 19km SE of Kodari, Nepal | 73-81 | 0-6 | 0.0 |
| 2015-06-04 | 6.0 | 116.54 | 5.99 | 14km WNW of Ranau, Malaysia | 66-89 | 4-11 | 0.0 |
| 2015-06-10 | 6.0 | -68.43 | -22.40 | 52km E of Calama, Chile | 5-12 | 6-31 | 0.0 |
| 2015-06-20 | 6.4 | -73.81 | -36.36 | 73km WNW of Talcahuano, Chile | 88-100 | 7-13 | 0.0 |
| 2015-07-03 | 6.1 | 125.89 | 10.17 | 23km NW of Santa Monica, Philippines | 79-81 | 14-18 | 0.0 |
| 2015-09-24 | 6.6 | 131.26 | -0.62 | 29km N of Sorong, Indonesia | 62-77 | 17-24 | 0.0 |
| 2015-11-07 | 6.8 | -71.45 | -30.88 | 39km SW of Ovalle, Chile | 56-81 | 0-6 | 0.0 |
| 2015-11-17 | 6.5 | 20.60 | 38.67 | 10km WSW of Nidri, Greece | 88-94 | 6-13 | 0.0 |
| 2015-12-17 | 6.6 | -93.63 | 15.80 | 12km SW of Tres Picos, Mexico | 58-78 | 11-26 | 0.0 |
| 2015-12-20 | 6.1 | 117.64 | 3.65 | 38km N of Tarakan, Indonesia | 66-74 | 6-11 | 0.0 |
| 2016-01-03 | 6.7 | 93.65 | 24.80 | 30km W of Imphal, India | 37-93 | 0-4 | 0.0 |
| 2016-01-11 | 6.2 | 141.09 | 44.48 | 74km NW of Rumoi, Japan | 68-100 | 2-17 | 0.0 |
| 2016-01-25 | 6.3 | -3.68 | 35.65 | 50km NNE of Al Hoceima, Morocco | 82-88 | 4-4 | 0.0 |
| 2016-02-10 | 6.3 | -71.58 | -30.57 | 36km W of Ovalle, Chile | 30-58 | 15-19 | 0.0 |
| 2016-04-13 | 6.0 | 122.02 | 7.79 | 15km NW of Siocon, Philippines | 55-70 | 0-11 | 0.0 |
| 2016-04-14 | 6.2 | 130.70 | 32.79 | 3km W of Kumamoto-shi, Japan | 57-100 | 2-13 | 0.0 |
| 2016-04-14 | 6.0 | 130.72 | 32.70 | 5km ENE of Uto, Japan | 29-100 | 6-24 | 0.0 |
| 2016-04-15 | 7.0 | 130.75 | 32.79 | 1km E of Kumamoto-shi, Japan | 24-35 | 7-20 | 0.0 |
| 2016-04-16 | 7.8 | -79.92 | 0.38 | 27km SSE of Muisne, Ecuador | 83-84 | 11-19 | 0.0 |
| 2016-04-20 | 6.2 | -80.21 | 0.64 | 20km WNW of Muisne, Ecuador | 94-94 | 0-4 | 0.0 |
| 2016-04-20 | 6.0 | -80.04 | 0.71 | 11km N of Muisne, Ecuador | 94-94 | 0-0 | 0.0 |
| 2016-04-25 | 6.0 | -93.15 | 14.48 | 83km WSW of Puerto Madero, Mexico | 23-89 | 0-0 | 0.0 |
| 2016-04-27 | 6.0 | -93.09 | 14.52 | 75km WSW of Puerto Madero, Mexico | 70-84 | 0-13 | 0.0 |
| 2016-05-31 | 6.4 | 122.55 | 25.56 | 94km ENE of Keelung, Taiwan | 70-84 | 2-15 | 0.0 |
| 2016-06-10 | 6.1 | -86.96 | 12.83 | 22km E of Puerto Morazan, Nicaragua | 83-89 | 0-11 | 0.0 |
| 2016-06-14 | 6.2 | 168.83 | -18.76 | 98km NNW of Isangel, Vanuatu | 84-94 | 9-22 | 0.0 |
| 2016-06-19 | 6.3 | 169.07 | -20.28 | 83km SSW of Isangel, Vanuatu | 89-94 | 0-9 | 0.0 |
| 2016-06-20 | 6.0 | 168.76 | -20.21 | 90km SW of Isangel, Vanuatu | 89-94 | 0-0 | 0.0 |
| 2016-07-20 | 6.1 | 169.05 | -18.93 | 72km NNW of Isangel, Vanuatu | 54-83 | 7-19 | 0.0 |
| 2016-08-24 | 6.2 | 13.19 | 42.72 | 10km SE of Norcia, Italy | 36-53 | 20-31 | 0.0 |
| 2016-09-17 | 6.0 | 140.57 | -2.08 | 51km NNW of Jayapura, Indonesia | 94-94 | 0-9 | 0.0 |
| 2016-10-26 | 6.1 | 13.07 | 42.96 | 3km NNW of Visso, Italy | 78-88 | 4-15 | 0.0 |
| 2016-11-08 | 6.0 | -73.56 | -36.58 | 42km WNW of Talcahuano, Chile | 82-94 | 9-20 | 0.0 |
| 2016-11-11 | 6.1 | 141.57 | 38.50 | 24km ENE of Ishinomaki, Japan | 93-100 | 2-9 | 0.0 |
| 2016-11-13 | 6.2 | 173.70 | -42.31 | 10km N of Kaikoura, New Zealand | 74-81 | 7-19 | 0.0 |
| 2016-11-13 | 6.1 | 173.62 | -42.18 | 26km N of Kaikoura, New Zealand | 74-81 | 7-19 | 0.0 |
| 2016-11-13 | 6.5 | 173.67 | -42.32 | 9km N of Kaikoura, New Zealand | 74-81 | 7-19 | 0.0 |
| 2016-11-13 | 7.8 | 173.05 | -42.74 | 54km NNE of Amberley, New Zealand | 74-81 | 7-19 | 0.0 |
| 2016-11-20 | 6.4 | -68.63 | -31.62 | 7km NNW of Pocito, Argentina | 14-52 | 26-46 | 0.0 |
| 2016-11-25 | 6.6 | 73.98 | 39.27 | 47km NE of Karakul, Tajikistan | 80-86 | 2-7 | 0.0 |
| 2017-02-18 | 6.4 | -66.66 | -23.86 | 52km NW of San Antonio de los Cobres, Argentina | 55-78 | 0-11 | 0.0 |
| 2017-04-05 | 6.1 | 60.44 | 35.78 | 61km NNW of Torbat-e Jam, Iran | 61-93 | 7-11 | 0.0 |
| 2017-05-12 | 6.2 | -90.06 | 12.92 | 78km SSW of Acajutla, El Salvador | 74-89 | 2-11 | 0.0 |
| 2017-05-20 | 6.0 | 123.95 | 9.38 | 25km S of Alburquerque, Philippines | 88-89 | 0-0 | 0.0 |
| 2017-05-29 | 6.6 | 120.43 | -1.29 | 28km WNW of Kasiguncu, Indonesia | 73-87 | 0-13 | 0.0 |
| 2017-06-14 | 6.9 | -92.01 | 14.91 | 2km SSW of San Pablo, Guatemala | 94-100 | 0-0 | 0.0 |
| 2017-06-22 | 6.8 | -90.97 | 13.72 | 28km SW of Puerto San Jose, Guatemala | 69-88 | 0-11 | 0.0 |
| 2017-07-20 | 6.6 | 27.41 | 36.92 | 11km ENE of Kos, Greece | 48-65 | 17-26 | 0.0 |
| 2017-09-19 | 7.1 | -98.40 | 18.58 | 5km ENE of Raboso, Mexico | 23-52 | 7-13 | 0.0 |
| 2017-09-19 | 7.1 | -98.49 | 18.55 | 1km ESE of Ayutla, Mexico | 23-52 | 7-13 | 0.0 |
| 2017-09-20 | 6.5 | 169.03 | -18.81 | 85km NNW of Isangel, Vanuatu | 47-82 | 6-11 | 0.0 |
| 2017-09-20 | 6.4 | 169.09 | -18.80 | 85km NNW of Isangel, Vanuatu | 47-82 | 6-11 | 0.0 |
| 2017-09-23 | 6.2 | -94.90 | 16.77 | 19km SE of Matias Romero, Mexico | 62-94 | 0-19 | 0.0 |
| 2017-09-23 | 6.1 | -94.95 | 16.74 | 18km SSE of Matias Romero, Mexico | 62-94 | 0-19 | 0.0 |
| 2017-09-23 | 6.1 | -94.95 | 16.75 | 17km SSE of Matias Romero, Mexico | 62-94 | 0-19 | 0.0 |
| 2017-10-10 | 6.3 | -69.64 | -18.52 | 36km SSW of Putre, Chile | 88-88 | 6-11 | 0.0 |
| 2017-10-18 | 6.0 | -173.89 | -20.60 | 100km SSE of Pangai, Tonga | 70-89 | 9-17 | 0.0 |
| 2017-11-12 | 7.2 | 45.79 | 34.93 | 32km SSW of Halabjah, Iraq | 20-38 | 15-33 | 0.0 |
| 2017-11-12 | 7.3 | 45.79 | 34.96 | 30km SW of Halabjah, Iraq | 20-38 | 15-33 | 0.0 |
| 2017-11-12 | 7.3 | 45.96 | 34.91 | 30km S of Halabjah, Iraq | 20-38 | 15-33 | 0.0 |
| 2017-11-12 | 7.3 | 45.94 | 34.89 | 32km S of Halabjah, Iraq | 20-38 | 15-33 | 0.0 |
| 2017-11-17 | 6.3 | 94.91 | 29.83 | 58km NE of Nyingchi, China | 25-55 | 4-4 | 0.0 |
| 2017-11-17 | 6.4 | 94.98 | 29.83 | 63km ENE of Nyingchi, China | 25-55 | 4-4 | 0.0 |
| 2017-11-17 | 6.4 | 94.98 | 29.83 | 63km ENE of Nyingchi, China | 25-55 | 4-4 | 0.0 |
| 2017-12-01 | 6.0 | 57.33 | 30.77 | 58km NNE of Kerman, Iran | 25-41 | 7-11 | 0.0 |
| 2017-12-12 | 6.0 | 57.27 | 30.84 | 64km NNE of Kerman, Iran | 16-27 | 0-11 | 0.0 |
| 2018-01-14 | 7.1 | -74.74 | -15.78 | 40km SSW of Acari, Peru | 54-65 | 0-0 | 0.0 |
| 2018-01-14 | 7.3 | -74.77 | -15.78 | 42km SSW of Acari, Peru | 54-65 | 0-0 | 0.0 |
| 2018-01-19 | 6.5 | -110.92 | 26.51 | 69km NE of Loreto, Mexico | 43-69 | 0-13 | 0.0 |
| 2018-01-19 | 6.3 | -111.11 | 26.68 | 77km NNE of Loreto, Mexico | 43-69 | 0-13 | 0.0 |
| 2018-01-21 | 6.3 | -69.62 | -18.89 | 76km S of Putre, Chile | 60-77 | 4-17 | 0.0 |
| 2018-01-24 | 6.2 | 142.29 | 41.12 | 92km ESE of Mutsu, Japan | 58-80 | 33-44 | 0.0 |
| 2018-02-04 | 6.1 | 121.68 | 24.16 | 21km NNE of Hualian, Taiwan | 80-84 | 7-7 | 0.0 |
| 2018-02-04 | 6.1 | 121.66 | 24.10 | 14km NNE of Hualian, Taiwan | 80-84 | 7-7 | 0.0 |
| 2018-02-06 | 6.4 | 121.66 | 24.13 | 18km NNE of Hualian, Taiwan | 56-72 | 4-7 | 0.0 |
| 2018-02-06 | 6.4 | 121.66 | 24.13 | 17km NNE of Hualian, Taiwan | 56-72 | 4-7 | 0.0 |
| 2018-02-06 | 6.4 | 121.65 | 24.17 | 22km NNE of Hualian, Taiwan | 56-72 | 4-7 | 0.0 |
| 2018-02-06 | 6.4 | 121.68 | 24.16 | 21km NNE of Hualian, Taiwan | 56-72 | 4-7 | 0.0 |
| 2018-06-21 | 6.1 | 168.02 | -17.82 | 33km WSW of Port-Vila, Vanuatu | 69-94 | 6-13 | 0.0 |
| 2018-06-21 | 6.1 | 168.06 | -17.80 | 28km WSW of Port-Vila, Vanuatu | 69-94 | 6-13 | 0.0 |
| 2018-06-30 | 6.0 | -105.09 | 19.07 | 44km WSW of San Patricio, Mexico | 74-84 | 0-15 | 0.0 |
| 2018-07-13 | 6.4 | 169.02 | -18.93 | 72km NNW of Isangel, Vanuatu | 74-88 | 0-9 | 0.0 |
| 2018-08-12 | 6.0 | -144.33 | 69.55 | 65km SSW of Kaktovik, Alaska | 86-93 | 11-19 | 0.0 |
| 2018-08-12 | 6.1 | -144.36 | 69.52 | 65km SSW of Kaktovik, Alaska | 86-93 | 11-19 | 0.0 |
| 2018-08-12 | 6.1 | -144.39 | 69.52 | 54km SSW of Kaktovik, Alaska | 86-93 | 11-19 | 0.0 |
| 2018-08-12 | 6.4 | -144.20 | 69.75 | 65km SSW of Kaktovik, Alaska | 86-93 | 11-19 | 0.0 |
| 2018-08-12 | 6.4 | -145.55 | 69.67 | 64km SW of Kaktovik, Alaska | 86-93 | 7-13 | 0.0 |
| 2018-08-12 | 6.1 | -145.25 | 69.62 | 64km SW of Kaktovik, Alaska | 86-93 | 7-13 | 0.0 |
| 2018-08-12 | 6.4 | -145.25 | 69.62 | 84km SW of Kaktovik, Alaska | 86-93 | 7-13 | 0.0 |
| 2018-08-12 | 6.3 | -145.30 | 69.56 | 90km SW of Kaktovik, Alaska | 86-93 | 7-13 | 0.0 |
| 2018-08-17 | 6.0 | -83.14 | 8.77 | 15km N of Golfito, Costa Rica | 84-94 | 0-9 | 0.0 |
| 2018-08-17 | 6.1 | -83.14 | 8.79 | 17km N of Golfito, Costa Rica | 84-94 | 0-9 | 0.0 |
| 2018-08-17 | 6.1 | -83.15 | 8.77 | 14km N of Golfito, Costa Rica | 84-94 | 0-9 | 0.0 |
| 2018-08-21 | 7.3 | -62.88 | 10.86 | 30km NE of Rio Caribe, Venezuela | 51-70 | 0-0 | 0.0 |
| 2018-08-21 | 7.3 | -62.91 | 10.74 | 20km NNW of Yaguaraparo, Venezuela | 51-70 | 0-0 | 0.0 |
| 2018-08-28 | 6.2 | 124.17 | -10.86 | 99km SE of Kupang, Indonesia | 78-83 | 11-13 | 0.0 |
| 2018-09-05 | 6.6 | 141.93 | 42.71 | 26km ESE of Chitose, Japan | 65-83 | 20-39 | 0.0 |
| 2018-09-05 | 6.7 | 142.00 | 42.70 | 31km ESE of Chitose, Japan | 65-83 | 20-39 | 0.0 |
| 2018-09-05 | 6.6 | 141.93 | 42.67 | 27km E of Tomakomai, Japan | 65-83 | 20-39 | 0.0 |
| 2018-09-07 | 6.2 | -78.90 | -2.35 | 17km SSW of Alausi, Ecuador | 69-69 | 4-4 | 0.0 |
| 2018-09-08 | 6.1 | 126.46 | 7.24 | 9km WNW of Manay, Philippines | 84-100 | 4-7 | 0.0 |
| 2018-10-25 | 6.8 | 20.56 | 37.51 | 33km SW of Mouzaki, Greece | 31-55 | 6-17 | 0.0 |
| 2018-10-25 | 6.8 | 20.56 | 37.48 | 35km SW of Lithakia, Greece | 31-55 | 6-17 | 0.0 |
| 2018-10-28 | 6.1 | -90.38 | 12.95 | 93km SW of Acajutla, El Salvador | 66-100 | 2-9 | 0.0 |
| 2018-10-30 | 6.1 | 174.98 | -39.05 | 64km ESE of Waitara, New Zealand | 79-95 | 7-13 | 0.0 |
| 2018-11-01 | 6.2 | -69.53 | -19.66 | 89km NE of Iquique, Chile | 68-77 | 13-24 | 0.0 |
| 2018-11-01 | 6.2 | -69.29 | -19.59 | 113km NE of Iquique, Chile | 68-77 | 13-24 | 0.0 |
| 2018-11-01 | 6.2 | -69.28 | -19.53 | 118km NE of Iquique, Chile | 68-77 | 13-24 | 0.0 |
| 2018-11-04 | 6.0 | 123.87 | 7.81 | 6km SE of Sapad, Philippines | 94-97 | 0-4 | 0.0 |
| 2018-11-10 | 6.2 | -173.82 | -20.54 | 98km SE of Pangai, Tonga | 89-100 | 7-22 | 0.0 |
| 2018-11-30 | 7.0 | -149.92 | 61.32 | 11km N of Anchorage, Alaska | 81-89 | 0-11 | 0.0 |
| 2018-11-30 | 7.0 | -149.92 | 61.32 | 13km N of Anchorage, Alaska | 81-89 | 0-11 | 0.0 |
| 2018-11-30 | 7.0 | -149.92 | 61.32 | 12km N of Anchorage, Alaska | 81-89 | 0-11 | 0.0 |
| 2018-11-30 | 7.0 | -149.94 | 61.34 | 13km N of Anchorage, Alaska | 81-89 | 0-11 | 0.0 |
| 2018-11-30 | 6.6 | -149.93 | 61.34 | 12km N of Anchorage, Alaska | 81-89 | 0-11 | 0.0 |
| 2019-01-05 | 6.8 | -71.57 | -8.14 | 89km W of Tarauaca, Brazil | 74-100 | 0-19 | 0.0 |
| 2019-01-08 | 6.3 | 131.04 | 30.59 | 16km SSE of Nishinoomote, Japan | 70-81 | 9-19 | 0.0 |
| 2019-01-18 | 6.0 | 168.63 | -19.21 | 76km WNW of Isangel, Vanuatu | 66-100 | 0-19 | 0.0 |
| 2019-02-01 | 6.6 | -92.30 | 14.76 | 14km ENE of Puerto Madero, Mexico | 49-69 | 4-20 | 0.0 |
| 2019-02-01 | 6.5 | -92.40 | 14.80 | 9km NNE of Puerto Madero, Mexico | 49-69 | 4-20 | 0.0 |
| 2019-02-22 | 7.7 | -76.90 | -2.30 | 132km ESE of Palora, Ecuador | 83-100 | 0-7 | 0.0 |
| 2019-02-22 | 7.5 | -77.02 | -2.20 | 115km ESE of Palora, Ecuador | 83-100 | 0-7 | 0.0 |
| 2019-03-01 | 7.0 | -70.13 | -14.68 | 27km NNE of Azangaro, Peru | 62-93 | 2-4 | 0.0 |
| 2019-03-08 | 6.0 | 126.08 | 10.39 | 40km N of Santa Monica, Philippines | 68-94 | 4-11 | 0.0 |
| 2019-03-15 | 6.3 | -65.89 | -17.86 | 28km S of Cliza, Bolivia | 52-94 | 0-37 | 0.0 |
| 2019-03-23 | 6.1 | -76.28 | 4.56 | 7km NW of El Dovio, Colombia | 73-93 | 0-7 | 0.0 |
| 2019-03-23 | 6.1 | -76.22 | 4.56 | 3km WSW of Versalles, Colombia | 73-93 | 0-7 | 0.0 |
| 2019-03-26 | 6.0 | -156.94 | 66.31 | 66km S of Kobuk, Alaska | 67-93 | 7-11 | 0.0 |
| 2019-03-31 | 6.2 | -80.81 | -1.99 | 27km N of Santa Elena, Ecuador | 89-89 | 6-7 | 0.0 |
| 2019-04-18 | 6.1 | 121.69 | 23.99 | 9km E of Hualian, Taiwan | 86-87 | 0-4 | 0.0 |
| 2019-04-18 | 6.0 | 121.69 | 24.07 | 13km NE of Hualian, Taiwan | 86-87 | 0-4 | 0.0 |
| 2019-05-06 | 7.1 | 146.40 | -6.98 | 37km NW of Bulolo, Papua New Guinea | 74-79 | 9-9 | 0.0 |
| 2019-05-06 | 7.2 | 146.44 | -6.98 | 33km NW of Bulolo, Papua New Guinea | 74-79 | 9-9 | 0.0 |
| 2019-05-09 | 6.1 | 131.85 | 31.78 | 43km ESE of Miyazaki-shi, Japan | 82-88 | 6-13 | 0.0 |
| 2019-05-09 | 6.1 | 131.85 | 31.78 | 42km ESE of Miyazaki-shi, Japan | 82-88 | 6-13 | 0.0 |
| 2019-05-09 | 6.3 | 131.80 | 31.80 | 37km ESE of Miyazaki-shi, Japan | 82-88 | 6-13 | 0.0 |
| 2019-05-30 | 6.6 | -89.27 | 13.24 | 27km SSE of La Libertad, El Salvador | 79-100 | 0-11 | 0.0 |
| 2019-06-18 | 6.4 | 139.45 | 38.64 | 33km WSW of Tsuruoka, Japan | 68-88 | 4-11 | 0.0 |
| 2019-06-18 | 6.4 | 139.47 | 38.65 | 31km WSW of Tsuruoka, Japan | 68-88 | 4-11 | 0.0 |
| 2019-09-29 | 6.7 | -73.16 | -35.48 | 69km WSW of Constitucion, Chile | 72-82 | 7-15 | 0.0 |
| 2019-10-16 | 6.4 | 125.00 | 6.71 | 7km ENE of Columbio, Philippines | 52-88 | 4-11 | 0.0 |
| 2019-10-16 | 6.7 | 125.13 | 6.78 | 5km W of Magsaysay, Philippines | 52-88 | 4-11 | 0.0 |
| 2019-10-21 | 6.4 | 169.45 | -19.04 | 59km NNE of Isangel, Vanuatu | 83-89 | 9-15 | 0.0 |
| 2019-11-27 | 6.0 | 23.27 | 35.73 | 41km NW of Platanos, Greece | 94-100 | 6-9 | 0.0 |
| 2019-12-15 | 6.8 | 125.21 | 6.72 | 5km SSE of Magsaysay, Philippines | 100-100 | 4-6 | 0.0 |
| 2019-12-15 | 6.8 | 125.19 | 6.71 | 6km S of Magsaysay, Philippines | 100-100 | 4-6 | 0.0 |
| 2019-12-24 | 6.0 | -74.04 | 3.50 | 3km SW of Lejanias, Colombia | 58-78 | 6-9 | 0.0 |
| 2019-12-24 | 6.0 | -74.00 | 3.49 | 4km SSE of Lejanias, Colombia | 58-78 | 6-9 | 0.0 |
| 2020-01-24 | 6.7 | 39.08 | 38.39 | 9km NNE of Doganyol, Turkey | 19-35 | 11-20 | 0.0 |
| 2020-01-24 | 6.7 | 39.08 | 38.33 | 4km ENE of Doganyol, Turkey | 19-35 | 11-20 | 0.0 |
| 2020-01-28 | 6.1 | -80.71 | 18.95 | 57km SE of East End, Cayman Islands | 62-89 | 0-13 | 0.0 |
| 2020-02-23 | 6.0 | 44.37 | 38.48 | 25km SE of Saray, Turkey | 70-80 | 4-9 | 0.0 |
| 2020-02-23 | 6.0 | 44.36 | 38.48 | 24km SE of Saray, Turkey | 70-80 | 4-9 | 0.0 |
| 2020-02-23 | 6.0 | 44.37 | 38.47 | 26km SE of Saray, Turkey | 70-80 | 4-9 | 0.0 |
| 2020-04-19 | 6.4 | 142.05 | 38.92 | 33km ESE of Ofunato, Japan | 51-76 | 4-30 | 0.0 |
| 2020-04-19 | 6.3 | 141.93 | 38.91 | 25km SE of Ofunato, Japan | 51-76 | 4-30 | 0.0 |
| 2020-05-02 | 6.6 | 25.71 | 34.20 | 89km S of Ierapetra, Greece | 54-72 | 6-19 | 0.0 |
| 2020-05-15 | 6.4 | -117.88 | 38.17 | 57km WNW of Tonopah, Nevada | 27-80 | 6-17 | 0.0 |
| 2020-05-27 | 6.1 | 168.10 | -18.73 | 112km SSW of Port-Vila, Vanuatu | 94-94 | 2-13 | 0.0 |
| 2020-05-27 | 6.1 | 167.86 | -17.16 | 80km NW of Port-Vila, Vanuatu | 94-94 | 2-13 | 0.0 |
| 2020-06-03 | 6.8 | -68.42 | -23.30 | 46km SSW of San Pedro de Atacama, Chile | 46-60 | 13-22 | 0.0 |
| 2020-06-03 | 6.8 | -68.50 | -23.40 | 60km SSW of San Pedro de Atacama, Chile | 46-60 | 13-22 | 0.0 |
| 2020-06-21 | 6.0 | -18.69 | 66.39 | 28km NNE of Siglufjordur, Iceland | 49-59 | 9-20 | 0.0 |
| 2020-06-23 | 7.4 | -95.91 | 16.04 | 20km NE of Santa Maria Xadani, Mexico | 58-79 | 0-0 | 0.0 |
| 2020-06-23 | 7.4 | -95.90 | 16.03 | 20km ENE of Santa Maria Xadani, Mexico | 58-79 | 0-0 | 0.0 |
| 2020-06-23 | 7.4 | -95.94 | 15.93 | 13km E of Santa Maria Xadani, Mexico | 58-79 | 0-0 | 0.0 |
| 2020-08-12 | 6.0 | 39.81 | -7.33 | 66 km ESE of Vikindu, Tanzania | 51-69 | 17-22 | 0.0 |

Table 1 The weather conditions 7 hours before the strong earthquake with magnitude of 6 or above. From left to right are the occurrence time and the magnitude of the earthquake, the longitude, the latitude, and the location of the epicenter, the relative humidity, wind speed and precipitation 7 hours before the strong earthquake.

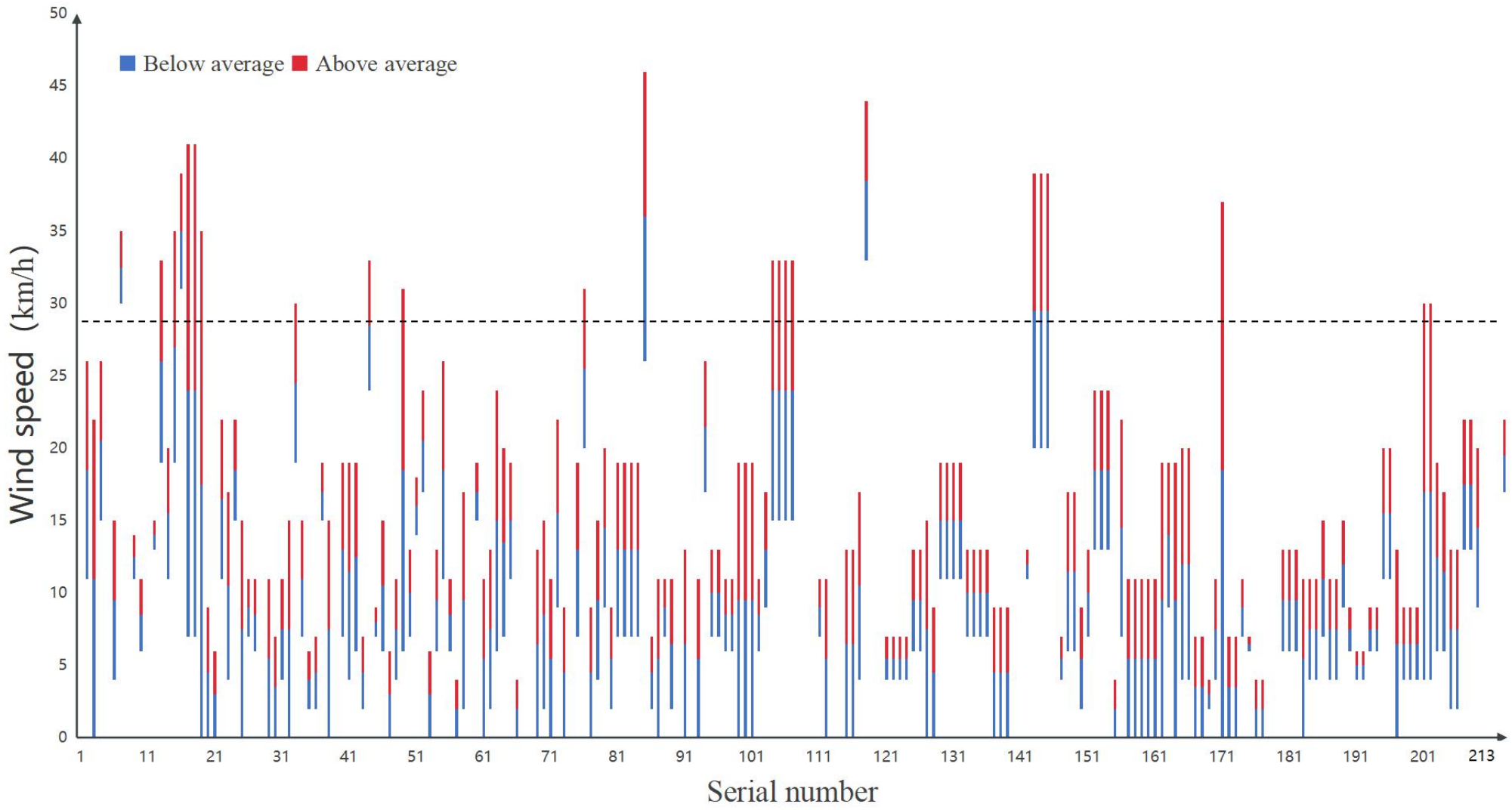

Fig.1 Graph of the statistical results of wind speed (each line represents the range of wind speed change in a statistical case, the junction of the red line and the blue line represents the average value of wind speed). The dashed line represents the wind speed is 28.8 km/h.

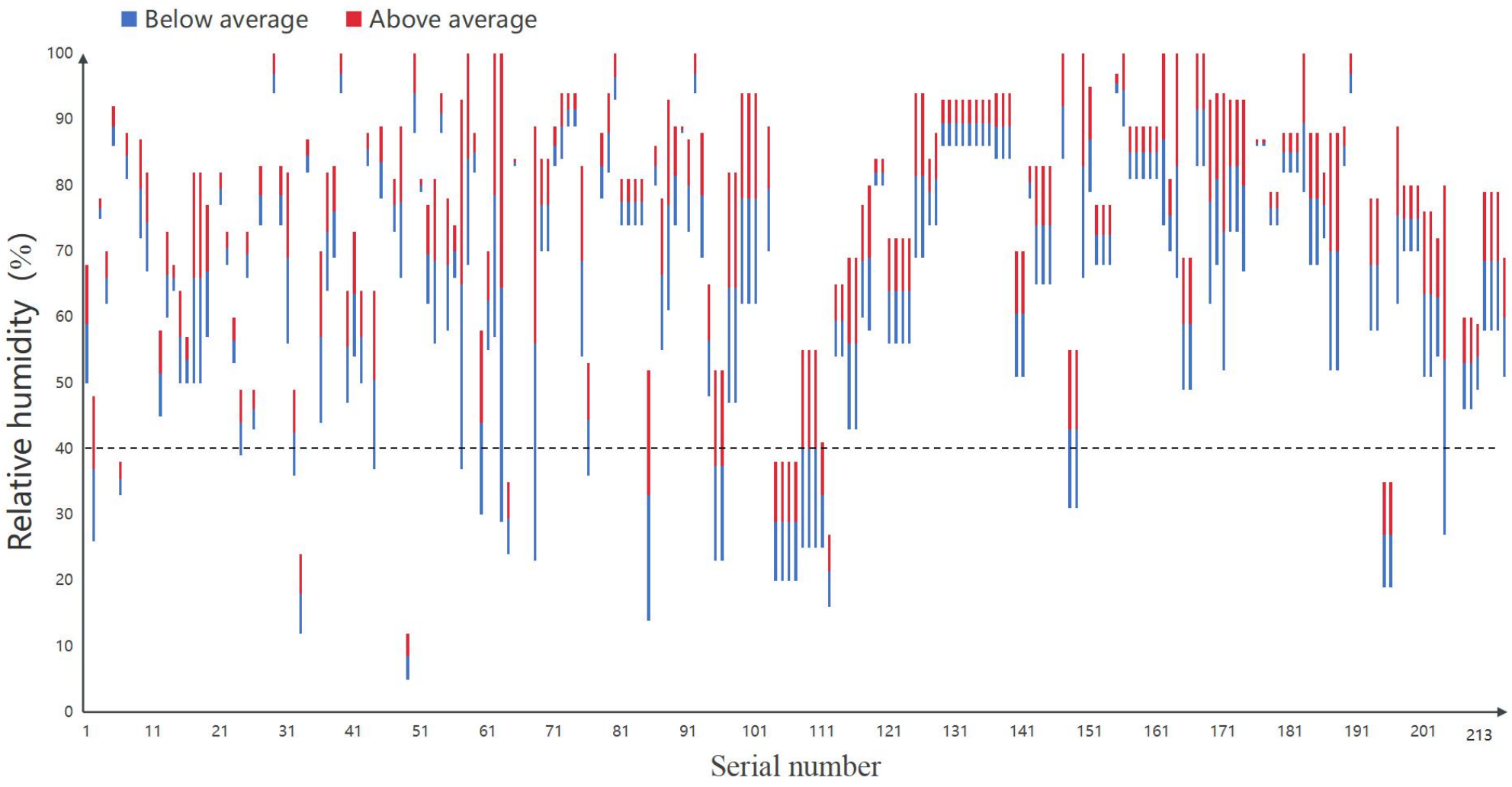

Fig.2 Graph of the statistical results of relative humidity (each line represents the range of relative humidity change in a statistical case, the junction of the red line and the blue line represents the average value of relative humidity). The dashed line represents the relative humidity is 40%.

The details of the strong earthquakes we selected are shown in Table 1. According to the statistical results of Table 1, there is no precipitation in all cases. For average wind speed as shown in Fig.1, only seven examples of them exceed 8 m/s (28.8 km/h) and maximum average wind speed is 10.69 m/s (38.5 km/h). For most cases, the average wind speed is less than 5.55 m/s (20 km/h). There were only a few cases with strong winds before strong earthquakes. However, most of them have high relative humidity, as shown in Fig.2, only 16 examples have an average relative humidity of less than 40%. Based on our definition of fair days, 96.7% of them are fair, despite their high relative humidity.

As for the distribution of seismogenic places, we have obtained 213 events of earthquakes in 36 different countries including Japan, Mexico, Canada, China, India and so on. Among them, there are 186 earthquake events of 7⩾M⩾6, accounting for 87.3% of the total events, and there are 26 earthquake cases of 8⩾M＞7, accounting for 12.2% of the total events. In addition, there is only one case with magnitude of more than 8, accounting for 0.5%.

## 4 Discussion and Conclusion

From all the cases we have studied in Section 3, 96.7% of the weather conditions 7 hours before earthquakes with magnitude of 6 or above are no precipitation with the average wind speed less than 8 m/s (28.8 km/h). It can be inferred that the weather before the strong earthquakes is (the statistics here are 6 class or above) almost be fair.

Why are the weather conditions fair 7 hours before the earthquake? Takeuchi et al. (2006) claimed that the electrostatic channel and the recombination of p-holes caused thermal anomalies before earthquakes (Tronin et al., 2002; Ouzonov and Freund, 2004; Liu et al., 1999). According to Mikhailova et al. (2018), cyclones can be the reason of the rising air temperature before earthquakes. Unlike these physical mechanisms, some research scholars have shown that the reason for the fair weather before the earthquake is more likely to be described below. The rocks will squeeze and collide with each other, and be hollowed out due to stress (Pulinets et al., 2005), and the soil will have cracks. A great number of studies on earthquakes were conducted that the radon concentration increased sharply before the strong earthquake (Holub and Brady, 1981; Segovia et al., 1989; King et al., 1993; Pulinets et al., 1997; Garavaglia et al. 2000; Yasuoka et al., 2006; Omori et al., 2007; Kawada et al., 2007). Based on this, it can be inferred that the plates squeeze each other to create micro-channels hereafter a large amount of radon gas is released and escaped through the cracks. Since radon gas is radioactive, a large amount of radon decay will generate a lot of energy. Part of the energy could heat the atmosphere. Liu et al. (1999) and Smirnov et al. (2017) also show that the temperature will rise before earthquakes. In general, this might be why there is no precipitation and fair 7 hours before the strong earthquake. Another part of the energy may be used to ionize the atmosphere, generating a large number of positive and negative charges. These positive and negative charges are separated, and a large number of positive charges gather near the ground to form a reverse electric field. This might give us the chance to observe the abnormal atmospheric electric field before the earthquake (Bonchkovsky, 1954; Kondo, 1968; Hao, 1988; Smirnov, 2008; Choudhury et al., 2013; Bychkov et al., 2017; Smirnov et al., 2017).

The atmospheric electric field is obviously affected by weather conditions before EQ compared to normal Ez under fair weather. So we could draw that "earthquakes of magnitude 6 or above are almost fair". If it is sunny and there is a continuous negative electric field anomaly, it is very likely that an earthquake will occur nearby within a few hours, which is important for early warning earthquakes. So, the mostly fair weather characteristics of continent EQ greater than Ms 6 is very useful to guaid strong EQ warning system in the future. Our work will provide a good prerequisite for earthquake early warning and eliminates the interference of weather conditions, which will be very helpful for earthquake early warning.

**Author contribution:** TC conceptualized the study. LL collected, processed and analyzed the data. LL and ST prepared the original draft with contributions from all authors. ZXX, MQM and WH were responsible for discussion and revision.

**Competing interests:** The authors declare that they have no conflict of interest.

**Acknowledgments:** The authors thank Prof. Fushan Luo, Prof. Jie Liu, Prof. Zhijun Niu and Prof. Shi Che for very useful discussion.

**Funding:** Supported by the Strategic Pioneer Program on Space Science, Chinese Academy of

Sciences, Grant No. XDA17010301, XDA17040505, XDA15052500, XDA15350201, and by the National Natural Science Foundation of China, Grant No. 41731070, 41931073. The authors thank some supports from the Specialized Research Fund for State Key Laboratories, and CAS-NSSC-135 project.